\documentclass[11pt]{article}
\usepackage{moriond,epsfig}

\def\be{\begin{equation}}
\def\ee{\end{equation}}
\def\bea{\begin{eqnarray}}
\def\eea{\end{eqnarray}}

\begin{document}
\title{SPIN CURRENT NOISE AS A PROBE OF INTERACTIONS}

\author{OLIVIER SAURET, \underline{DENIS FEINBERG}}

\address{LEPES, CNRS, BP 166, 38042 Grenoble, FRANCE}

\maketitle\abstracts{
 The spin resolved 
current shot noise can uniquely probe the interactions in mesoscopic systems: i) in a  
normal-superconducting junction,
 the spin current noise is zero, as carried by singlets, 
and ii) in a single electron 
transistor (SET) in the sequential regime, the spin current noise is Poissonian. 
Coulomb interactions 
lead to usually repulsive, but also attractive
correlations. Spin current shot noise 
can also be used to measure the spin relaxation 
time $T_1$.}

Non-equilibrium (shot) noise provides information about the charge 
and the statistics of carriers in mesoscopic systems \cite{BB}. The Pauli exclusion principle
 leads to a reduction of shot noise from the Schottky value \cite{MartinLandauer}. 
Coulomb interactions also act in correlating 
wavepackets, yet the Coulomb interactions may decrease or increase noise correlations \cite{coulomb}. 
Thus, in a given mesoscopic structure, the effects on the shot noise 
of Fermi statistics and of interactions are intimately mixed. 
In contrast, we propose here that spin-resolved shot noise 
can unambiguously probe the effects of interactions \cite{prl}. In a nutshell,
 the Pauli principle acting only on electrons with the same spin, currents 
wavepackets carried by quasiparticles with opposite spins can only be correlated 
by the interactions. "Spin current noise" has received little attention before, and
 with a different purpose. For instance, spin shot noise was recently considered in 
absence of charge current \cite{wang}, and the effect of a spin-polarized current on 
charge and spin noise was investigated
\cite{Bulka}. Noise is also an efficient probe for testing quantum 
correlations in two-electron spin-entangled states 
\cite{entangle}.

In contrast, let us consider mesoscopic structures in 
which the average current is {\it not} 
spin-polarized, but where the currents carried by quasiparticles with different 
spins can be separately measured. First, consider a 
 mesoscopic device made of a normal metal with non-interacting electrons, 
non magnetic terminals $i,j$. In absence of 
magnetic fields and spin scattering , the scattering matrix is spin-independent, 
 $s_{ij}^{\sigma\sigma'}=\delta_{\sigma\sigma'}s_{ij}$. Then one verifies that
the spin-resolved noise, defined as 
 $S_{ij}^{\sigma\sigma'}(t-t')=
 \frac{1}{2}\langle 
 \Delta I_i^{\sigma}(t)\Delta I_j^{\sigma'}(t')+\Delta 
 I_j^{\sigma'}(t')\Delta I_i^{\sigma}(t) \rangle$
 where $\Delta 
 I_i^{\sigma}(t)=I_i^{\sigma}(t)-\langle I_i^{\sigma} \rangle$, is diagonal in the spin variables, 
 $S_{ij}^{\sigma\sigma'}(\omega)=\delta_{\sigma\sigma'}S_{ij}(\omega)$. 
 Thus, choosing an arbitrary spin axis $\bf z$, the total (charge) current 
 noise $S_{ij}^{ch} = S_{ij}^{\uparrow \uparrow} + S_{ij}^{\downarrow 
\downarrow}
 + S_{ij}^{\uparrow \downarrow} + S_{ij}^{\downarrow \uparrow}$ and the 
 {\it spin current noise} $S_{ij}^{sp} = S_{ij}^{\uparrow \uparrow} + 
S_{ij}^{\downarrow \downarrow}
 - S_{ij}^{\uparrow \downarrow} - S_{ij}^{\downarrow \uparrow}$, defined 
 as the correlation of the {\it spin currents}
$I_{i}^{sp}(t)=I_i^{\uparrow}(t)-I_i^{\downarrow}(t)$,  
 are strictly equal. On the contrary, in presence of interactions, 
one expects that $S_{ij}^{\uparrow 
 \downarrow}= S_{ij}^{\downarrow \uparrow} \neq 0$, or equivalently 
$S_{ij}^{sp} \neq S_{ij}^{ch}$. 
 
Let us first consider a NS junction, where S is a singlet superconductor 
and N a normal metal. The scattering 
matrix coupling electron (e) and holes (h) 
in the metal is made of spin-conserving normal terms $s_{ee}^{\sigma 
\sigma}$, $s_{hh}^{\sigma \sigma}$, and Andreev terms
$s_{eh}^{\sigma -\sigma}$, $s_{he}^{\sigma -\sigma}$ coupling opposite 
spins. The total zero-
frequency noise $S^{ch}=\sum_{\sigma \sigma'}S^{\sigma \sigma'}$ is given at
zero temperature by the
well-known result\cite{NS} $S^{ch}=\frac{4e^3V}{\pi \hbar} 
Tr[s_{he}^{\dagger}s_{he}(1-s_{he}^{\dagger}s_{he})]$.  
We have in turn calculated the spin-resolved correlations 
$S^{\sigma \sigma}$ and $S^{\sigma -\sigma}$, 
and found that they are exactly {\it equal}. 
As a result, for a NS junction, at $T = 0$, the 
spin current shot noise 
is strictly zero, $S^{sp}=0$. The current correlation between electrons 
with opposite spins is $S^{\uparrow \downarrow}=S^{\uparrow \uparrow}$,
 therefore {\it positive}. This "bunching" of opposite spins carriers is 
an obvious consequence of the Andreev process, e. g. the transmission 
of singlets through the interface. It has been recently discussed 
in a three-terminal geometry\cite{FazioSFF}. 
  
\begin{figure}[thpb]
\centerline{\epsfig{file=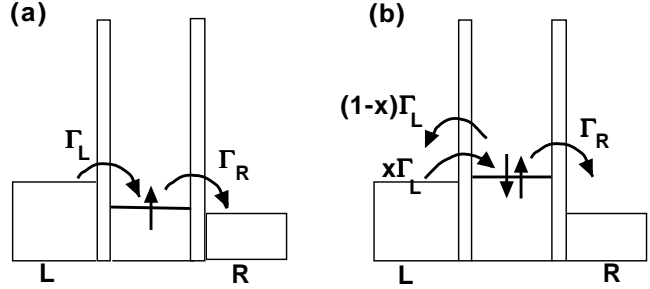,width=9cm,angle=0}} 
\caption{The SET transport sequence a) Between charge 
states $N=0$ and $1$ : rates 
$\Gamma_L$ and $\Gamma_R$; b) Between 
charge 
states $N=1$ and $2$ : rates 
$x\Gamma_L$ from reservoir $L$, $(1-x)\Gamma_L$ to reservoir $L$ and 
$\Gamma_R$ 
to reservoir $R$.} 
\label{fig:SET}
\end{figure}

Let us now consider a small quantum dot in 
the sequential transport regime, where repulsive correlations are instead
expected. It is connected by tunnel 
 barriers to normal leads $L$ and $R$ with potentials 
 $\mu_{L,R}$, with $eV=\mu_{L}-\mu_{R}$ (Fig. \ref{fig:SET}).
 One assumes that max$(eV,k_BT) >> \hbar 
 \Gamma_{L,R}$ and that only one level of energy $E_0$ sits between 
$\mu_{R}$ 
 and $\mu_{L}$. The dot can be in three possible occupation states 
($N=0,1,2$) of the level (Fig. \ref{fig:SET}). 
$U(N)$ being the Coulomb energy 
for the state $N$, $\Delta E_{L,R}^{+}(N)=E_0-\mu_{L,R} + U(N+1)-U(N)$ is needed
 to add an electron to state $N$ from 
leads $L,R$, and $\Delta E_{L,R}^{-}(N)=-E_0+\mu_{L,R} + U(N-1)-U(N)$ is needed
 to remove an electron from state 
$N$ towards $L,R$. Let us further assume that $\Delta E_{L}^{+}(0)$, 
$\Delta E_{R}^{-}(1)<< -k_BT$, which implies that
the transitions from $N=0$ to $1$ involve electrons coming only from $L$, 
and the transitions from $N=1$ to $0$ involve 
electrons going only into $R$. One allows the Coulomb energy to vary and 
consider the 
possibility of transitions from $N=1$ to $2$, only from $L$, e.g. 
$\Delta E_{R}^{-}(2)<< -k_BT$. This describes the following situation : if  
$\Delta E_{L}^{+}(1) >> k_BT$, the transition to state $N=2$ is forbidden 
and only two charge states $N=0$, $1$ are involved (Fig. \ref{fig:SET}a). 
If on the contrary $\Delta E_{L}^{+}(1) << -k_BT$, then the three charge 
states $0$, $1$, $2$ are involved (Fig. \ref{fig:SET}b).
This physical situation corresponds for instance to 
fixing the gate voltage such as $U(1)=U(0)$, and varying the 
ratio between $k_BT$ and the Coulomb excess energy $U(2)-U(1)$.   

Let us write the master equation describing this system 
\cite{prl}. Assuming a constant density of 
states in the reservoirs and 
defining $x$ as the Fermi function $x = [1+exp(\beta \Delta 
E_{L}^{+}(1))]^{-1}$, the populations $p_0$, $p_{\uparrow}$, 
$p_{\downarrow}$ and $p_2$ verify

\begin{eqnarray}
\label{eq:master}
 \nonumber
 &\dot{p}_0 = -2\Gamma_L\, p_0 + \Gamma_R \,(p_{\uparrow} + 
p_{\downarrow})\\
 \nonumber
 &\dot{p}_{\uparrow} = -(\Gamma_R + x\Gamma_L) \,p_{\uparrow} + \Gamma_L\, 
p_0 
 + ((1-x)\Gamma_L +\Gamma_R)\, p_2\\
 &\dot{p}_{\downarrow} = -(\Gamma_R + x\Gamma_L)\, p_{\downarrow} + 
 \Gamma_L\, p_0 
 + ((1-x)\Gamma_L +\Gamma_R)\, p_2\\
\nonumber
 &\dot{p}_2 = -2((1-x)\Gamma_L +\Gamma_R)\, p_2 + 
 x\Gamma_L \,(p_{\uparrow} + p_{\downarrow})
 \end{eqnarray}

Let us first consider the limit $x=1$, corresponding to a resonant state 
without charging energy. Then spin 
 $\uparrow$ and $\downarrow$ currents are uncorrelated, the average current is 
$\langle I \rangle = 2e\frac{\Gamma_L\Gamma_R}{\Gamma_L + 
 \Gamma_R}$, the total zero-frequency noise \cite{ChengTing} $S_{ij}(\omega = 
0)=2e\langle I 
 \rangle(1-\frac{2\Gamma_L\Gamma_R}{(\Gamma_L + 
 \Gamma_R)^2})$. Here 
$S_{ij}^{\uparrow 
 \downarrow}= S_{ij}^{\downarrow \uparrow} = 0$, or equivalently 
$S^{sp}=S^{ch}$. This is another 
example of uncorrelated transport.
 
 Let us now consider the SET case $x=0$, where charge transport is 
maximally correlated. The charge noise is given by $S_{ij}(\omega = 0)=2e\langle I 
 \rangle(1-\frac{4\Gamma_L\Gamma_R}{(2\Gamma_L + 
 \Gamma_R)^2})$ \cite{nazarovstruben}. Apart from an effective doubling of 
the rate $\Gamma_L$, this 
 result is qualitatively similar to that obtained without interactions. 
Therefore the charge noise is not the best possible 
probe of interactions. 
 On the contrary, the behaviour of the spin noise is 
completely different. Using the method by Korotkov 
\cite{korotkov}, we find that
 
\begin{eqnarray}
&S_{ij}^{\sigma \sigma}= e\langle I \rangle 
(1-\frac{2\Gamma_L\Gamma_R}{(2\Gamma_L + \Gamma_R)^2}),\;\;\;
S_{ij}^{\sigma -\sigma}= -e\langle I \rangle 
\frac{2\Gamma_L\Gamma_R}{(2\Gamma_L + \Gamma_R)^2},\;\;\;
&S_{ij}^{sp}=2e\langle I \rangle
\label{eq:bruitSET}
\end{eqnarray}

 \noindent
The result for $S^{sp}$ resembles a 
 Poisson result (maximal fluctuations). 
 The correlations between currents of opposite spins are 
 negative, like a partition noise. Yet spin-up and 
 spin-down channels are separated 
 as wavepackets with up or down spins exclude 
 each other because of interactions, rather than statistics.
Here, each junction is -- due to Coulomb repulsion -- sequentially crossed
by elementary wavepackets with well-defined but uncorrelated 
spins. On the contrary, {\it  charge} current
wavepackets are correlated on times $\sim \hbar/\Gamma_i$, leading to the 
reduction as compared to the Poisson value. 
Notice that the analysis of the SET involving $N=1$ and $2$ states (instead of 
$0$, $1$) yields exactly the same result. 
 
 The general solution of Eqs. (\ref{eq:master}) spans the full regime 
between the uncorrelated and the maximally correlated cases. 
The average current is given by $\langle I 
\rangle=e\frac{2\Gamma_L \Gamma_R}{\Gamma_R + (2-x) \Gamma_L}$. The spin 
current noise 
components $S_{ij}^{\sigma \sigma'}$ (i,j=L,R) can also be calculated. The
expression for the spin noise is $
S_{ij}^{sp}=2e\langle I \rangle \,(1-\frac{2x\Gamma_L \Gamma_R}{(\Gamma_R 
+ \Gamma_L)(\Gamma_R+x\Gamma_L)})$.

\begin{figure}[thpb]
\centerline{\epsfig{file=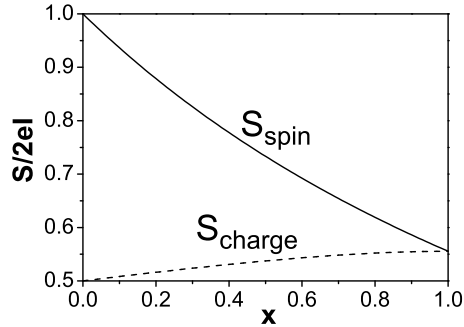,width=7cm,angle=0}} 
\caption{Spin shot noise and charge shot noise in the 
SET, as a function of $x$ 
(see text) : $x=0$ denotes the maximal correlation, $x=1$ the uncorrelated 
case.
$\Gamma_R=2\Gamma_L$ : antibunching of opposite spins.} 
\label{fig:bruit2}
\end{figure}

\begin{figure}[thpb]
\centerline{\epsfig{file=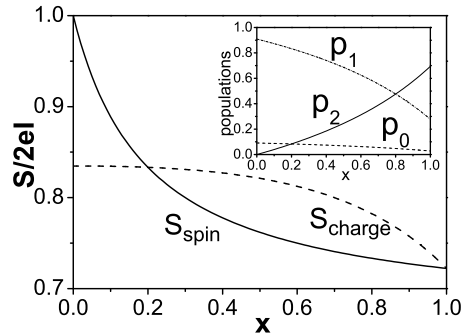,width=7cm,angle=0}} 
\caption{Same as Fig. 2, $\Gamma_R=0.2\Gamma_L$ :
bunching of opposite spins for $x>x_c$. The inset shows the probabilities 
of states $N=0,1,2$ and the population inversion at large $x$.}
\label{fig:bruit0.2}
\end{figure}

The expression for the total (charge) noise $S^{ch}$ is too lengthy to be 
written here. Figs.
\ref{fig:bruit2}, \ref{fig:bruit0.2} 
show the variation with $x$ of the
charge and spin current noise. The spin noise is maximum for $x=0$, 
decreases monotonously and merges the charge noise at $x=1$. 
The role of the asymmetry of the junctions is very striking. First, if 
$\Gamma_R > \Gamma_L$, $S^{sp}$ is always larger 
than $S^{ch}$ (Fig. \ref{fig:bruit2}), like in the ideal SET ($x=0$). 
On the contrary, if $\Gamma_R < \Gamma_L$, $S^{sp}$ is
smaller than $S^{ch}$ for $x > x_c \sim \Gamma_R/\Gamma_L$ (Fig. 
\ref{fig:bruit0.2}). This implies that $S^{\uparrow \downarrow} > 0$, 
contrarily to the 
naive expectation for repulsive interactions : if $\Gamma_R < \Gamma_L$, 
the low charge states are unfavored and the high ones favored, despite of 
Coulomb repulsion. Two electrons tend to enter the dot 
successively, with opposite spins, leading to a certain degree of 
bunching. Here the anomaly is due to a kind of 
"population inversion",  
manifesting a strong departure 
from equilibrium (Fig. \ref{fig:bruit0.2}). 

\begin{figure}[thpb]
\centerline{\epsfig{file=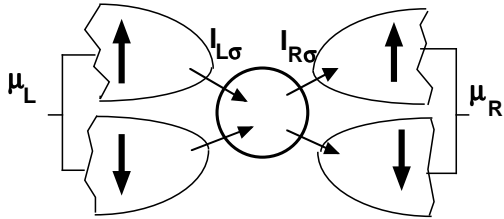,width=9cm,angle=0}}
\caption{Schematic set-up for spin current measurement, 
using four spin-polarized terminals (see text).}
\label{fig:4term}
\end{figure}

Including a spin-flip rate $T_1^{-1}= \gamma_{sf}$, 
one finds $S_{LR}^{sp}=2e \langle I 
\rangle \,\frac{\Gamma_R}{\Gamma_R + \gamma_{sf}}$, 
which suggests\cite{prl} a method to measure $T_1$.
Fig. \ref{fig:4term} shows a possible four-terminal set-up \cite{spinflip} 
for the measurement of spin current 
correlations, with ferromagnetic leads. 
In a fully symmetric 
device, the net current flowing through the 
SET is not spin 
polarized. Yet it is in principle possible to
measure the noise correlations
$S_{L1L1}$, $S_{L1L2}$, $S_{L1R1}$, $S_{L1R2}$, etc... If each terminal 
generates a fully spin-polarized current,
the analysis of this set-up can be mapped onto the above model. 
If polarization 
is not perfect, the
above measurement should mix spin noise with charge 
noise. If those are sufficiently different 
(strong repulsive correlations), they could still be distinguished, 
allowing to probe the Coulomb correlations
 by the method of spin current noise.

\end{document}